\documentclass[a4paper]{jpconf}
\usepackage{graphicx}
\usepackage{amssymb}
\usepackage{epsfig}
\newcommand{\real}{{\mathbb R}} %% real numbers
\def\e{{\,\rm e}\,}

\newcommand{\proj}{{\mathbb P}}

\def\ii{{\,{\rm i}\,}}
\def\dd{{\rm d}}

\def\beq{\begin{equation}}
\def\bee{\begin{equation}}
\def\eeq{\end{equation}}
\def\bea{\begin{eqnarray}}
\def\eea{\end{eqnarray}}
\def\bd{\begin{displaymath}}
\def\ed{\end{displaymath}}
\begin{document}
\begin{flushright}
 HWM-05-30 \\
  EMPG-05-21
\end{flushright}
\title{Black-holes, topological strings and large N phase transitions}

\author{N. Caporaso$^{(a)}$, M. Cirafici$^{(b)}$, L. Griguolo$^{(c)}$, S. Pasquetti$^{(c)}$, D. Seminara$^{(a)}$
and  R. J. Szabo$^{(b)}$ }

\address{$^{(a)}$ Dipartimento di Fisica, Polo Scientifico Universit\`a di
Firenze, INFN Sezione di Firenze Via  G. Sansone 1, 50019 Sesto
Fiorentino, Italy\\
$^{(b)}$ Department of Mathematics and
Maxwell Institute for Mathematical Sciences, Heriot-Watt University,
Colin Maclaurin Building, Riccarton, Edinburgh EH14 4AS, UK\\
$^{(c)}$ Dipartimento di  Fisica, Universit\`a  di Parma, INFN
Gruppo Collegato di Parma, Parco Area delle Scienze 7/A, 43100
Parma, Italy}

%\ead{caporaso@fi.infn.it, M.Cirafici@ma.hw.ac.uk,
%griguolo@fis.unipr.it, pasquetti@fis.unipr.it, seminara@fi.infn.it, R.J.Szabo@ma.hw.ac.uk}

\begin{abstract}
The counting of microstates of BPS black-holes on local Calabi-Yau
of the form ${\mathcal O}(p-2)\oplus{\mathcal
O}(-p)~\longrightarrow~S^2$ is explored   by computing the partition
function of $q$-deformed Yang-Mills theory on $S^2$. We obtain, at
finite $N$, the instanton expansion of the gauge theory. It can be
written exactly as the partition function for $U(N)$ Chern-Simons
gauge theory on a Lens space, summed over all non-trivial vacua,
plus a tower of non-perturbative instanton contributions.  In the
large $N$ limit we find a peculiar phase structure in the model. At
weak string coupling the theory reduces  to the trivial  sector and
the topological string partition function on the resolved conifold
is reproduced in this regime. At a certain critical point,
instantons are enhanced and the theory  undergoes a phase transition
into a strong-coupling regime. The transition from the
strong-coupling phase to the weak-coupling phase is of third order.
\end{abstract}
%\vskip -20cm
\section{Introduction}
A novel and highly non trivial relation between the topological string vacuum amplitude
$Z_{\rm top}$ and the partition function $Z_{\rm BH}$ of  $N=2$ BPS black-holes
in four dimensions has been recently  suggested in
\cite{Ooguri:2004zv}: $Z_{\rm BH}$=$|Z_{\rm
top}|^2$. This intriguing equality
generalizes a series of beautiful results
\cite{LopesCardoso:1998wt,Mohaupt:2000mj}
concerning the entropy of BPS black holes arising in
compactifications of Type~II superstrings on Calabi-Yau threefolds and
it is supposed to hold for a large
black hole charge $N$ at any order in
the $\frac1N$ expansion, once taken into account the {\it perturbative}
definition of $Z_{\rm top}$.  Actually the proposal of \cite{Ooguri:2004zv}
is even more startling, since it suggests to employ the above relation as
a non perturbative definition of the topological string dynamics.

\noindent
In order to check this proposal, one should engineer  Calabi-Yau
backgrounds in which both sides of the relation can be computed
independently. For this task, in \cite{Aganagic:2004js}, generalizing the original example presented
in \cite{Vafa:2004qa},  the following  class of  non-compact
Calabi-Yau threefolds,
\beq {\cal M}={\mathcal O}(p+2g-2)\oplus{\mathcal
O}(-p)~\longrightarrow~\Sigma_g \ , \label{totalhol}\eeq
has been considered. The manifold $\Sigma_g$ is a Riemann
surface of genus $g$, while
${\mathcal O}(m)$ is a holomorphic line bundle of degree $m$ over
$\Sigma_g$.
The counting of BPS states on these geometries has been claimed to
reduce to computing the partition function of a peculiar deformation
of Yang-Mills theory on $\Sigma_g$ called $q$-deformed Yang-Mills.
 Starting from this result one should ask if the relation
with the perturbative topological string amplitudes holds in this
case. Happily the partition function $Z_{\rm top}$ for these
geometries has been computed very recently \cite{Bryan:2004iq} and
the consistency check amounts to reproducing these amplitudes as the
large $N$ limit of $q$-deformed Yang-Mills theory on $\Sigma_g$. In
\cite{Aganagic:2004js} a large $N$ expansion has been performed and
the conjecture was confirmed, but with a couple of important
subtleties. Firstly, one should include in the definition of $Z_{\rm
top}$ a sum over a $U(1)$ degree of freedom identified with a
Ramond-Ramond flux through the Riemann surface. Secondly, and more
importantly, the relevant topological string partition function
implies the presence of $|2g-2|$ stacks of D-branes inserted in the
fibers of ${\cal M}$. An explanation of this unexpected feature in terms of
extra closed string moduli, related to the non-compactness of ${\cal M}$,
has been offered in \cite{Aganagic:2005dh}. Recent works on this
topic include \cite{Dabholkar:2004yr}--\cite{Parvizi:2005aa}.

 That
two-dimensional Yang-Mills theory should be related to a string
theory in the large $N$ limit is not entirely unexpected due to the
well-known Gross-Taylor expansion
\cite{Gross:1992tu}--\cite{Gross:1994mr}. At large $N$, the
partition function of two-dimensional Yang-Mills theory on a Riemann
surface $\Sigma_g$ almost factorizes into two copies (called chiral
and antichiral) of the same theory of unfolded branched covering
maps, the target space being $\Sigma_g$ itself. The
chiral-antichiral factorization is, however, violated by some geometrical
structures called orientation-reversing tubes.

The emergence of chiral and antichiral sectors was also observed in
\cite{Aganagic:2004js} in studying the $q$-deformed version of two
dimensional Yang-Mills theory and it implies the appearance of the
modulus squared $|Z_{\rm top}|^2$, a crucial ingredient in checking
the relation with black hole physics. However, on $S^2$, this
picture may be jeopardized  by the $q$-deformed incarnation of the
Douglas-Kazakov phase transition occurring at large $N$. There, a
strong-coupling phase, wherein the theory admits the Gross-Taylor
string description, is separated by a weak-coupling phase with
gaussian field theoretical behaviour. Instanton configurations
induce the transition to strong-coupling \cite{Gross:1994mr}, while
the entropy of branch points appears to be responsible for the
divergence of the string expansion above the critical
point~\cite{Taylor:1994zm,Crescimanno:1994eg}.

Here we shall explore in detail $q$-deformed Yang-Mills theory
on $S^2$ elucidating its relation with topological string theory on the
threefold ${\cal M}={\mathcal O}(p-2)\oplus{\mathcal
O}(-p)\to\mathbb{P}^1$ and its rich phase diagram. The details of this
analysis can be found in \cite{Caporaso:2005ta} and related work in \cite{Arsiwalla:2005jb,Jafferis:2005jd}.

\section{Black holes, topological strings and $ \mathbf{q}$-deformed Yang-Mills
  theory}

\subsection{The conjecture}
The conjecture presented in \cite{Ooguri:2004zv} is more easily
phrased in the context of  Type~IIA superstring theory compactified
on $\mathcal{M}\times\real^{3,1}$, where $\mathcal{M}$ is a
Calabi-Yau threefold. In this framework a BPS black hole can be
realized  by wrapping D6, D4, D2 and D0 branes around holomorphic
cycles in $\mathcal{M}$.  The  D2 and D0-brane charges are referred
to as  ``electric'',  while D6 and D4-brane charges are
``magnetic''. One can define a partition function for a mixed
ensemble of BPS black hole states by fixing the magnetic charges
$Q_6$ and $Q_4$ and summing over the D2 and D0 charges with fixed
chemical potentials $\phi_2$ and $\phi_0$ to get
\begin{equation}
Z_{\rm
BH}(Q_6,Q_4,\phi_2,\phi_0)=\sum_{Q_2,Q_0}\Omega(Q_6,Q_4,Q_2,Q_0)
~\exp\bigl[-Q_2\phi_2-Q_0\phi_0\bigr] \ , \label{ZBH}
\end{equation}
where $\Omega(Q_6,Q_4,Q_2,Q_0)$ is the \textit{number} of  BPS
states with fixed D-brane charges. The conjecture relates the
partition function (\ref{ZBH}) to the  (A-model) topological string
partition function   $Z_{\rm top}(g_s,t_s)$ on $\cal{M}$
\begin{equation}Z_{\rm BH}(Q_6,Q_4,\phi_2,\phi_0)=\bigl|Z_{\rm
    top}(g_s,t_s)\bigr|^2 \ ,
\end{equation}
where the topological string coupling $g_s$ and the K\"ahler modulus
$t_s$ are related to the black-hole charges as follows
\begin{equation}
g_s=\frac{4\pi \ii}{\frac{\ii}{\pi}\,\phi_0+Q_6} \ ,
\,\,\,\,\,\,\,\,\,\,t_s=
\mbox{$\frac{1}{2}$}\,g_s\bigl(\mbox{$-\frac{\ii}{\pi}$}\, \phi_2+N
Q_4\bigr)
\end{equation}
%\cite{Neitzke:2004ni,Vonk:2005yv}
The origin of this  proposal can be traced back to some well-known
properties of BPS black holes in ${\cal N}=2$
supergravity~\cite{LopesCardoso:1998wt,Mohaupt:2000mj}.
These are solutions interpolating between two maximally
supersymmetric vacua: the Minkowski space at infinity and the
Bertotti geometry in the near horizon region. Moreover they  carry
charges $(P^I,Q_I)$ with the respect to $n_v+1$ abelian gauge fields
present in the theory: the $n_v$ matter vector multiplets and the
graviphoton. The entropy for this class of geometries is
determined by the Bertotti mass $M_{\rm Bert}$ appearing in the near
horizon geometry and, in turn, $M_{\rm Bert}$ is a function of the
scalar fields $X^I$ of the vector multiplets at the horizon.  This
potential dependence  on the scalars is problematic, because it contrasts with the
idea that the black-hole  entropy is a universal quantity, fixed by
the conserved charges (\textit{no-hair theorem}). What comes to
rescue is the so-called \textit{attractor
mechanism}~\cite{Ferrara:1995ih,Strominger:1996kf}.
%Black hole solutions are found in
%the background of $2 n_v+2$ gauge fields ($n_v+1$ magnetic duals of
%the others) and they carry charges $(P^I,Q_I)$ with respect to them.
%The gauge fields are organized into $n_v$ vector multiplets plus one
%graviphoton (arising from the supergravity multiplet).
The scalar fields of the theory $X^I$%, that may be regarded as
%position dependent moduli of the Calabi-Yau threefold on which
%superstrings have been compactified,
have fixed values at the black
hole horizon determined only by the charges, through the geometric
equations\footnote{To be precise, this supergravity analysis is phrased in the mirror setting of
type IIB.}
\begin{equation}P^I={\rm Re}(X^I)=\oint_{A_I}{\rm Re}(\Omega) \
\ \ \  \ \ \mathrm{and}\  \  \  \ \  Q_I={\rm Re}(F_I)=\oint_{B^I}{\rm Re}(\Omega). \
\label{Peri}
\end{equation}
Here $\Omega$  is the holomorphic three-form on the Calabi-Yau $\mathcal{M}$
while  $(A_I,B^I)$ are a basis of $H_3(\mathcal{M},\mathbb{Z})$.
The period  $F_I$ is the gradient of  the prepotential $F_0(=X^I F_I)$.
By means of the
explicit expressions in eq.~(\ref{Peri}) one obtains
$ S_{\rm BH}\simeq Q_I X^I-P^IF_I \ .
$ The entropy thus appears as a Legendre transform
of the prepotential $F_0$, which coincides with the genus zero
free energy of topological strings in the Calabi-Yau background. At
the supergravity level, one can include higher-derivative
corrections proportional to $R^2T^{2g-2}$, where $T$ is the
graviphoton field-strength, and  recompute the black hole solutions
and their entropies
\cite{LopesCardoso:1998wt,Mohaupt:2000mj}.
At genus $g=1$, the relation with the
topological string free energy still holds when one includes quantum
corrections to the prepotential coming from one-loop amplitudes
\cite{Antoniadis:1993ze,Bershadsky:1993cx}. The suggestion of
\cite{Ooguri:2004zv} is  obviously consistent with these results and
cleverly suggests a generalization of them to all orders in the perturbative
string expansion.
%Actually, the proposal  of \cite{Ooguri:2004zv} is even  more
%suggestive. In fact,  eq.~(\ref{ZBH}) is proposed as  a
%nonperturbative definition of $Z_{\rm top}$, since   $Z_{\rm BH}$ makes sense for any value of the couplings.
%In particular the presence of a square-modulus signals a
%breaking of holomorphicity at the nonperturbative level.
It is of course natural to attempt to check this conjecture on some
explicit  Calabi-Yau threefold $\mathcal{M}$. While for the compact case the
task seems out of reach presently, in the non-compact case there is
the general class of threefolds (\ref{totalhol}) on which the
problem has been attacked \cite{Aganagic:2004js,Vafa:2004qa}. The
study of topological strings on these backgrounds and the related
counting of microstates have also produced a number of interesting
mathematical results. As we will see, different gauge theories in
diverse dimensions appear to be related by their common
gravitational ancestor.

\subsection{Counting microstates in ${\cal N}=4$ and $q$-deformed
Yang-Mills theories}

Let us begin by describing the counting of microstates. It consists
of counting bound states of D4, D2 and D0-branes, where $N$ D4
branes wrap the four-cycle $C_4$, which is the total space of the
holomorphic line bundle ${\mathcal O}(-p)~\longrightarrow~\Sigma_g $,
and the D2-branes wrap the Riemann surface $\Sigma_g$. The natural way of doing the computation is by
studying the relevant gauge theory on the brane. According to the general
framework \cite{Bershadsky:1995qy} the worldvolume gauge theory on
the $N$ D4-branes is the ${\cal N}=4$ topologically twisted $U(N)$
Yang-Mills theory on $C_4$.  The presence
of chemical potentials for D2 and D0-branes  is
simulated by turning on the observables in the theory given by
\begin{equation} \label{4daction}
S_{c}=\frac{1}{2 g_s}\,\int_{C_4}{\rm Tr}\bigl(F\wedge
F\bigr)+\frac{\theta}{g_s}\,\int_{C_4}{\rm Tr}\bigl(F\wedge K\bigr)
\ ,
\end{equation}
where $F$ is the Yang-Mills field strength and $K$ is the unit
volume form of $\Sigma_g$. The relation between the gauge parameters
$g_s$, $\theta$ and the chemical potentials are $\displaystyle{
\phi_0={4\pi^2}/{g_s}\ \ \mathrm{and}\ \  \phi_2={2\pi\,
\theta}/{g_s}} $
in accordance with the identifications for the D0 and D2-brane
charges $q_0$, $q_2$ as
\begin{equation}
q_0=\frac{1}{8\pi^2}\,\int_{C_4}{\rm Tr}\bigl(F\wedge F\bigr) \ ,
\,\,\,\,\,\,q_2=\frac{1}{2\pi}\,\int_{C_4}{\rm Tr}\bigl(F\wedge
K\bigr) \ .
\end{equation}
Evaluating $Z_{\rm BH}$ is thus  equivalent to computing the
expectation value in topologically twisted ${\cal N}=4$ Yang-Mills
theory given by
\begin{equation}
Z_{\rm BH}=\Bigl\langle\exp\Bigl[-\frac{1}{2 g_s}\,\int_{C_4}{\rm
Tr}\bigl(F\wedge F\bigr)-\frac{\theta}{g_s}\,\int_{C_4}{\rm
Tr}\bigl(F\wedge K\bigr)\Bigr]\Bigr\rangle=Z_{{\cal N}=4} \ .
\end{equation}
The general structure of this functional integral has been explored
in~\cite{Vafa:1994tf}. There it was shown that the partition
function $Z_{{\cal
    N}=4}$ has an expansion of the form
\begin{equation}
Z_{{\cal N}=4}=\sum_{q_0,q_2}\Omega(q_0,q_2;N)~
\exp\Bigl(-\frac{4\pi^2}{g_s}\,q_0-\frac{2\pi\,\theta}{g_s}\,q_2\Bigr)
\ , \label{N4}
\end{equation}
where $\Omega(q_0,q_2;N)$ is  the Euler
characteristic of the moduli space of $U(N)$ instantons on $C_4$ in
the topological sector labelled by the zeroth and second Chern
numbers $q_0$ and $q_2$. The counting of microstates is therefore
equivalent to an instanton counting in the ${\cal N}=4$ topological
gauge theory. This is still a formidable problem, because
no general strategy exists in the case of non-compact manifolds and
very few results \cite{Bruzzo:2002xf,Fucito:2004ry,nakajima} have
been obtained in this context.

However in~\cite{Vafa:2004qa}   the
computation was related to  a two-dimensional problem. In fact by introducing
certain massive deformations and  under the reasonable assumption that
the path integral localizes to $U(1)$-invariant modes around the fiber ${\cal O}(-p)$,
it was  argued  that the theory reduces to an effective
gauge theory on $\Sigma_g$.
In \cite{Vafa:2004qa} it was also shown that the non-triviality of the
line bundle ${\cal O}(-p)$ generates an extra term in the effective
two-dimensional action of the form \beq
S_p=-\frac{p}{2g_s}\,\int_{\Sigma_g}{\rm Tr}\,\Phi^2~K
\eeq
where $\Phi(z)$ parameterizes
the holonomy of the gauge field $A$ around a circle at infinity in
the fiber over the point $z\in\Sigma_g$. The relevant two-dimensional action
becomes
\begin{equation}
S_{{\rm YM}_2}=\frac{1}{g_s}\,\int_{\Sigma_g}{\rm Tr}\bigl(\Phi\,
F\bigr)+\frac{\theta}{g_s}\,\int_{\Sigma_g}{\rm Tr}\,\Phi~K
-\frac{p}{2g_s}\,\int_{\Sigma_g}{\rm Tr}\,\Phi^2 K.\label{aqd}
\end{equation}
%the first two terms being obtained by effective  dimensional
%reduction.
This is just the action of two-dimensional Yang-Mills theory on the
Riemann surface $\Sigma_g$. However there is an important subtlety. The
new degree of freedom $\Phi$  is periodic due to its origin as the
holonomy of the gauge field at infinity and this
 periodicity  affects the path integral
measure  in a well-defined way \cite{Aganagic:2004js}.  The final result  is
\begin{equation}
Z_{\mathcal{N}=4}=Z^{q}_{\rm YM}\!\!=\!\!\!\sum_{R}{\rm
dim}_q(R)^{2-2g}q^{\frac{p}{2}\,C_2(R)}\e^{\ii\theta\,
C_1(R)}\ \  \mathrm{with}\ \ {\rm dim}_q(R)=\!\!\!\!\!\!\prod_{1\leq i<j\leq
N}\!\!\!\!\!\!\frac{[R_i-R_j+j-i]_q}{[j-i]_q}. \label{pqd}
\end{equation}
This is to be compared with the partition function of ordinary
Yang-Mills theory on $\Sigma_g$ given by the Migdal expansion
\cite{Migdal:1975zg} \begin{equation} Z_{\rm YM}=\sum_{R}{\rm
dim}(R)^{2-2g}~\e^{\frac{g^2A}{2}\,C_2(R)}~\e^{\ii\theta\, C_1(R)}
\ \ \ \ \mathrm{with}\ \ \ \ \ {\rm dim}(R)=\!\!\!\!\!\!\prod_{1\leq i<j\leq N}\!\!\!\!\!\frac{R_i-R_j+j-i}{j-i}
,\label{migdal}
\end{equation}
where $R$ runs through the unitary irreducible representations of
the gauge group $U(N)$, ${\rm dim}(R)$ is its dimension, and
$C_1(R)$ and $C_2(R)$ are respectively its first and second Casimir
invariants. The dimensionless combination $g^2A$ of the Yang-Mills
coupling constant $g^2$ and area $A$ of $\Sigma_g$ is the effective
coupling of the theory, as dictated by invariance under
area-preserving diffeomorphisms. The main effect of the deformation of
the path integral measure  is to turn
the ordinary dimension  ${\rm dim} R$
into the {\it quantum} dimension ${\rm dim}_q R$.
The deformation parameter $q$ is related to the coupling $g_s$
through $q=\e^{-g_s}$ and the deformed integer number $[n]_q\equiv (q^{n/2}-q^{-n/2})/(q^{1/2}-q^{-1/2})$ .  Clearly as $g_s\to 0$ the quantum
dimension goes smoothly into the classical one. The effective
dimensionless coupling of the gauge theory is $g^2A=g_sp$.
% and
%therefore the topological character of the theory is preserved, as
%there is no effective dependence on the intrinsic area of
%$\Sigma_g$.
%and therefore the topological character of the theory is
%preserved, as there is no effective dependence on the intrinsic area
%of $\Sigma_g$.

Eq.~(\ref{pqd}) provides the solution of the difficult problem of
instanton counting, as it represents the answer on all
four-manifolds $C_4$. The direct comparison with eq. (\ref{N4})
seems at first sight puzzling, because the expansion of the
$q$-deformed gauge theory is in $\e^{-g_s}$ and not in
$\e^{-{1}/{g_s}}$. This implies that a modular transformation is
required in eq.~(\ref{pqd}) and we will discuss this point in
Sect.~3.

\subsection{Large $N$ limit and topological strings}
 Eq.~(\ref{pqd}) gives, in
principle, the exact expression for the black hole partition
function $Z_{\rm BH}$. In order to establish the connection with the
perturbative topological string partition function on $\mathcal{M}$, we need
to take the limit of large charges, in other words the large $N$ limit
of the $q$-deformed Yang-Mills theory. It is important, first of
all, to set the relation between the parameters of the gauge theory
describing the black hole and the data of the closed topological
string theory.
For the K\"ahler modulus $t_s$ of the base $\Sigma_g$, with the help
of the attractor mechanism, one finds
\begin{equation}
\label{15}
t_s= 2\pi \ii \frac{X^1}{X^0}=(p+2g-2)\,\frac{Ng_s}{2}-\ii\theta \ .
\end{equation}
The closed topological string emerging from
the large $N$ limit is thereby expected to possess the modulus (\ref{15}).
 This is
the first very non-trivial feature that should be reproduced by the
large $N$ limit.
The second non-trivial point is the  square modulus structure. This is not completely unexpected due to the relation
with the large $N$ limit of ordinary Yang-Mills theory. In the
Gross-Taylor expansion~\cite{Gross:1993yt} two types of
representations contribute to the limit, called the chiral
representations (with much less than $N$ Young tableaux boxes) and
the antichiral representations (with order $N$ boxes). The partition
function is almost factorized into two copies (apart from subtleties).
One could wonder whether
$Z^q_{\rm YM}$ factorizes as well. The choice of relevant
representations is dictated by the Casimir dependence of the
partition function, which is unchanged by the deformation. We expect
$
Z^q_{\rm YM}\simeq Z^{q,+}_{\rm YM}\,Z^{q,-}_{\rm YM} \ .
$
Moreover, one would expect that the chiral $q$-deformed Yang-Mills
partition function $Z^{q,+}_{\rm YM}$ could be written as a
holomorphic function of $t_s$ and identified with the topological
string amplitude $Z_{\rm top}(g_s,t_s)$ on $\mathcal{M}$. In
\cite{Aganagic:2004js} these expectations have been confirmed, but
with some  subtleties as well.

\noindent
For $g=0$, the case studied extensively in this proceeding, the result is
\begin{equation}
Z^q_{\rm YM}(S^2)=\sum_{l=-\infty}^{\infty}~\sum_{ \hat R_1, \, \hat
R_{2}}Z^{q{\rm YM},+}_{ \hat R_1, \, \hat R_2}(t_s+p\,g_s l)\,
Z^{q{\rm YM},-}_{ \hat R_1, \, \hat R_2}(\,\bar{t}_s-p\,g_s l)
\end{equation}
with $ Z^{q{\rm YM},-}_{ \hat R_1, \, \hat
R_2}(\,\bar{t}_s\,)=(-1)^{| \hat R_1|+| \hat R_2|}\, Z^{q{\rm
YM},+}_{ \hat R_1^\top, \, \hat R_2^\top}(\,\bar{t}_s\,) .
$
 Here $\hat R_{i}$ are irreducible representations of $SU(N)$ and
the chiral block $Z^{q{\rm YM},+}_{ \hat R_1, \hat
R_{2}}(t_s)$ agrees exactly with the perturbative topological
string amplitude on ${\cal M}$ \cite{Bryan:2004iq} with $2$ stacks of
D-branes inserted in the fiber.
Here $| \hat R|$ is the total number of boxes of the Young tableau
of the representation $ \hat R$. The chiral and anti-chiral
parts are sewn along the D-branes and summed over them. The extra
sum over the integer $l$ originates from the
$U(1)$ degrees of freedom contained in the original gauge group
$U(N)$. The generalization to higher genus
follows the same path, but it is slightly different \cite{Aganagic:2004js}.

 The genus zero case is also special
because it admits a standard description in terms of toric geometry.
The fibration $X={\cal O}(p-2)\oplus {\cal O}(-p)\to\proj^1$ is in
fact a toric manifold \cite{CGPS} and the partition function can be
written in terms of the topological vertex  $C_{ \hat R_1 \hat R_2
\hat R_3}(q)$ \cite{Aganagic:2003db} as
\begin{equation}
Z^{q{\rm YM},+}_{ \hat R_1, \, \hat R_2}(t_s)=Z_0~ q^{{k_{ \hat
R_1}}/{2}}~\e^{-\frac{t_s(| \hat R_1|+| \hat R_2|)}{p-2}}~
\sum_{\hat R}\e^{-t_s| \hat R|}~q^{{(p-1)k_{ \hat R_1}}/{2}}~C_{0
\hat R_1 \hat R^\top}(q)\,C_{0 \hat R \hat R_2}(q) \ ,
\end{equation}
where $k_{\hat R}$ is related to the Young tableaux labels through
$k_{\hat R}=\sum_i \hat R_i( \hat R_i-2 \, i+1)$ and $Z_0$
represents the contribution from constant maps. This is the
partition function of the topological A-model on $\mathcal{M}$ with
non-compact D-branes inserted at two of the four lines in the web
diagram. The D-branes are placed at a well-defined ``distance"
$\hat t=t_s/(p-2)$ from the Riemann surface, thereby introducing another
geometrical modulus.

We thus observe an apparent discrepancy between the prediction of
\cite{Ooguri:2004zv} that $Z_{\rm BH}=|Z_{\rm top}|^2$ and the
explicit computation leading to $Z_{\rm
BH}=\sum_{b,l}\,\bigl|Z^{(b,l)}_{\rm top}\bigr|^2 , $ where index $b$ denotes
the sum over chiral blocks with branes inserted. However, the extra sum
over the integer $l$  has been naturally interpreted as a sum over R-R
fluxes through $\Sigma_g$ \cite{Vafa:2004qa}. The sum over
the fiber D-branes is instead related to the fact that a
non-compact Calabi-Yau   may have additional moduli coming from the
non-compact directions \cite{Aganagic:2005dh}. This ``external"
sum is weighted with a different K\"ahler parameter
$\hat{t}=t_s/(p-2)$, and the partition function
effectively depends on two parameters. This
suggests that $\hat{t}$ could be interpreted as a new
K\"ahler modulus  \cite{CGPS}.

The whole picture therefore seems very convincing and pointing
towards a beautiful confirmation of the conjecture presented in
\cite{Ooguri:2004zv}.  However, there is a point that has been overlooked
which could have interesting ramifications. In taking the large $N$
limit it is possible to encounter phase transitions. The prototype
of this kind of phenomenon was discovered long ago
\cite{Gross:1980he,Wadia:1980cp} in the one-plaquette model of
lattice gauge theory.

A well-known large $N$ phase transition is the Douglas-Kazakov
transition~\cite{Douglas:1993ii}. It concerns Yang-Mills theory on
the sphere, a close relative of the relevant black hole ensembles
discussed earlier. A strong-coupling phase, in which the theory
admits the Gross-Taylor string description with its
chiral-antichiral behaviour, is separated by a weak-coupling phase
with gaussian field theoretical behavior. Two-dimensional Yang-Mills
theory on $S^2$ is equivalent to a string theory only above a
certain critical value of the effective t'Hooft coupling constant
$\lambda=Ng^2A$ given by $\lambda_c=\pi^2
.$ Instanton configurations induce the phase transition
to the strong-coupling regime \cite{Gross:1994mr}. On the other
hand, the entropy associated to certain classes of branched covering
maps seems responsible for the divergence of the string perturbation
series above the critical point
\cite{Taylor:1994zm,Crescimanno:1994eg}. It is natural to expect
that, if $q$-deformed large $N$ Yang-Mills theory is related to the
undeformed one, some of these features could find a place in the
black holes/topological string scenario. %Note that the partition
%function $Z_{BH}$ describing the black-hole physics at large charges
%is { not} the topological string partition function, but rather its
%square modulus summed over a complicated set of boundary conditions.
%The final result is thus { not} an analytic function of the K\"ahler
%parameters, preventing in principle its analytical continuation
%below certain points.
We will explore this possibility in the
subsequent sections.

\section{Large $N$-limit of $q$-deformed $YM_2$: one-cut solution and weak-coupling regime}

We discuss now the large $N$ limit of $q$-deformed Yang-Mills
theory on the sphere. An explicit result for the leading order
(planar) contribution to the free energy of ordinary Yang-Mills
theory on the sphere was obtained in \cite{Douglas:1993ii}. For
large area it fits nicely with the interpretation in terms of
branched coverings that arises in the Gross-Taylor expansion, down
to the phase transition point at $\lambda_c=\pi^2$ where the
string series is divergent. We will now perform similar
computations for $q$-deformed Yang-Mills theory on $S^2$. We start
by defining the relevant parameters to be held fixed as
$N\to\infty$ by
$ t=g_s N$ and by $a=g_s p N=p\,t  $.
The partition function of the $q$-deformed gauge theory on $S^2$
is given by \beq \label{cas}{Z}_{\rm YM}^{q}(g_s,p)= \sum_{n_1, \dots, n_N\in
\mathbb{Z}\atop n_i-n_j \ge i-j \ \mathrm{for}\ i\ge j}
\e^{-\frac{g_s p}{2}\, {( n_1^2+\cdots+n_N^2)}}\,\prod_{1\le
i<j\le N} \sinh^2\left(\mbox{$\frac{g_s}{2}(n_i-n_j)$}\right), \
\label{ZYMqnewvars} \eeq where we have chosen $\theta=0$.
The constraint on the sums keeps track of
the meaning of the integers $n_i$ in terms of Young tableaux
labels and highest weights. In terms of these new variables, the
partition function at $N\to\infty$ is given by $ {Z}_{\rm
YM}^{q}(t,a)=\exp\bigl(N^2 S_{\rm eff}(\rho)\bigr) $ where \beq
\label{pippino4} S_{\rm eff}=-\int_{C} \dd w~ \int_{C} \dd
w^\prime \rho(w)\,\rho(w^\prime\,)\,
 \log\left[\sinh\left(\mbox{$\frac{t}{2}$}\,|w-w^\prime|\right)\right]+
\frac{a}{2 }\,\int_C \dd w ~\rho(w) \ w^2 \ . \eeq We have
introduced the variable $x_i=i/N$ and the function $n(x)$ such
that $n(x_i)=\frac{n_i}{N}$. In the large $N$ limit, $x_i$ becomes
a continuous  variable $x\in[0,1]$: the density $\rho$ is defined
as $\rho(n)=\frac{\partial x(n)}{\partial n}$ and we denoted the
interval $[n(0),n(1)]$ by $C$. The distribution $\rho(z)$ in
eq.~(\ref{pippino4}) can be determined by requiring that it
minimizes the action. This implies that it satisfies the
saddle-point equation \beq \label{cippino5a} \frac{a}{2}\,
z=\frac{t}{2}\,\int_C \dd w ~\rho(w)\, \coth\left(\mbox{$
\frac{t}{2}$} \,(z-w)\right) \ . \eeq This equation is a
deformation of the usual Douglas-Kazakov equation that governs
ordinary QCD$_2$ on the sphere. The ordinary gauge theory is
recovered when $t\to 0$ while $a$ is kept fixed. The one-cut solution of
eq.~(\ref{cippino5a}) is given in \cite{Marino:2004eq} and  its explicit
form is given by \beq \rho(z)=
\frac{a }{\pi\, t }\, \arctan\left(\mbox{$\sqrt{\frac{
\e^{t^2/a}}{\cosh^2(\frac{t \,z}{2})}-1}$}~\right) \eeq with the
symmetric support $
z\in\bigl[-\mbox{$\frac{2}{t}$}\, \mathrm{arccosh}(\e^{-t^2/2
a})~,~\mbox{$\frac{2}{t}$}\, \mathrm{arccosh}(\e^{-t^2/2 a})
\bigr] \ .$ It can be considered as the $q$-deformation of the
well-known Wigner semi-circle distribution. We now come to the
crucial point: in the continuum limit the constraint on the series
(\ref{ZYMqnewvars}) becomes $n(x)-n(y)\ge x-y$ for $x\ge y$: we
may translate the original constraint in terms of the function
$\rho$ as $\rho(n)\le 1 $ \cite{Douglas:1993ii}.
The above bound  is of
fundamental importance. In fact its violation signals a
 potential large $N$ phase transition with the consequent existence of a
strong-coupling phase.  In terms of the variables $t$ and
$p$, the  bound on $\rho$ produces the inequality\beq \arctan\left(\sqrt{{
\e^{t/p}}-1}~\right)\le \frac{\pi}{p} \ , \label{pippino6}\eeq a
condition that is always satisfied for $p=1$ or $p=2$. The
situation changes for $p>2$ and the inequality (\ref{pippino6})
can be equivalently written as $\sqrt{{ \e^{t/p}}-1}\le
\tan\frac{\pi}{p}$ which implies that \beq t\le t_c=
p\,\log\left(\sec^2\left(\mbox{$\frac{\pi}{p}$}\right)\right)  \ .
\eeq Our solution, therefore, breaks down when the '{t}~Hooft
coupling $t$ reaches the critical value $t_c$: we remark that the
cases $p=1,2$ are special because then the one-cut
solution is always valid. The breakdown of the one-cut solution
parallels exactly what happens in ordinary two-dimensional
Yang-Mills theory, where it signals the appearance of a phase
transition. In the saddle-point approach it is possible to go
further and to find a solution describing the strong-coupling
phase, for $t>t_c$. Before proceeding with this analysis, we
compute the free energy in the weak-coupling phase and we discuss
its topological string interpretation.
% In our case we can still
%use the relation \beq \frac{\partial\mathcal{F}( t,a)}{\partial
%a}=\frac{1}{2}\,\int_{-b}^b \dd x~x^2\, \rho(x) \ , \eeq as in the
%usual one-matrix model with a gaussian potential.
A tedious but
elementary integration gives \beq {\cal F}(t,a)=-\frac{t^2}{6 a}
+\frac{\pi^2 a}{6 t^2} - \frac{a^2}{t^4}\,\zeta(3) +
\frac{a^2}{t^4}~\mathrm{Li}_3(\e^{- {t^2}/{a} })+c(t) \ .
\label{conif}\eeq The $a$-independent function $c(t)$ can be
easily determined by looking at the asymptotic expansion in $t$.
In the limit $t\to 0$, $p\to \infty$ the free-energy of ordinary
Yang-Mills theory in the weak-coupling phase is of course
recovered. We observe that the free energy depends only on the
combination $t^2/a$ (or $t_s/(p(p-2))$ in string variable). This
dependence is in contrast with the one expected from the geometry
of the Calabi-Yau ${\mathcal O}(p-2)\oplus{\mathcal
O}(-p)~\longrightarrow~S^2 $. This picture, in fact, contains
as a necessary ingredient two independent moduli $e^{-t_s}$ and
$e^{-\frac{2 t_s}{p-2}}$: we miss the Calabi-Yau geometry (there
is no substantial dependence on $p$, that simply scales the only
K\"ahler modulus without affecting the structure of the
free-energy) and the modulus square. It appears clear that in the
weak-coupling regime we cannot reproduce the geometrical structure
predicted by the conjecture: while for $p>2$ we could expect that
above the critical point a strong-coupling solution will do it, we
conclude that this is impossible for $p=1,2$. Actually the
free-energy eq.(\ref{conif}) coincides with the genus 0
free-energy of closed topological string theory on the resolved
conifold. Thus in the weak-coupling phase we have $Z_{\rm
qYM}$=$Z_{\rm top}^{Res.}$ with K\"ahler modulus $t^2/a$. The loss
of the original geometric information encoded in $q$-deformed
$YM_2$ and the mysterious appearance of the resolved conifold can
be easily understood by studying the large $N$ limit of the gauge
theory in the dual instanton picture.

%\section{Instanton driven large $\mathbf{N}$ phase transition}

The geometrical meaning of the $q$-deformed theory on $S^2$ becomes
more transparent when we consider the dual description in terms of
instantons, provided by a modular transformation of the series
(\ref{cas}). This is accomplished by means of a Poisson resummation.
It is also an efficient way to investigate the behaviour of the
theory at weak-coupling $g_s$, as we shall explain soon. By
exploiting the properties of the Stieltjes-Wigert orthogonal
polynomials, we can easily obtain the instanton expansion of
$q$-deformed Yang-Mills theory on $S^2$ \beq \label{cipow}Z_{\rm
YM}^q(g_s,p)=\frac{1}{N!}\, \sum_{s_i\in
\mathbb{Z}}\e^{-\frac{2\pi^2 }{g_s p}\,\sum_{i=1}^N s_i^2}
~{w}^{\mathrm{inst}}_q(s_1,\dots,s_N) \  \eeq where \bea
\label{cipow2}
{w}_q^{\mathrm{inst}}(s_1,\dots,s_N)\!\!&\!\!\!\!\!=\!\!\!\!\!&\!\!\frac{1}{2}\,\left(\frac{2\pi}{g_s
p}\right)^N \!\!\!\!\!\!\e^{-\frac{{g_s}\left(  N^3-N \right) }{6
p}}\,\int^\infty_{-\infty}\!\!\!\! \dd z_1\cdots \dd z_N
~\e^{-\frac{2\pi^2}{g_s p}\,
\sum_{i=1}^N  z_i^2}\nonumber\\
&&\times\,\prod_{1\le i< j\le N}
  \left[ \cos \left(\mbox{$\frac{2\,\pi \,\left( {s_i} - {s_j} \right)
        }{p}$}\right) - \cos
  \left(\mbox{$\frac{2\,\pi \,\left( {z_i} - {z_j} \right)
      }{p}$}\right) \right] \ .
\eea Our terminology mimicks that of the undeformed theory where the
partition function can be computed exactly via a nonabelian localization
\cite{Witten:1992xu}. It is given by a
sum over contributions localized at the classical solutions of the
theory. For finite $N$ the $U(N)$ path integral is given by a sum
over unstable instantons where each instanton contribution is
given by a finite, but non-trivial, perturbative expansion. By
``instantons'' we mean solutions of the classical Yang-Mills field
equations, which are not
gauge transformations of the trivial solution $A=0$. On the sphere
$S^2$, the most general solution  is
given by $ \bigl(A(z)\bigr)_{ij}=\delta_{ij}~A^{(m_i)}(z)$ where
$A^{(m_i)}(z)$ is the Dirac monopole potential of magnetic charge
$m_i$. The Yang-Mills action evaluated  on such an instanton is
given by $S_{\mathrm{\rm inst}}=\frac{2\pi^2}{g^2A}\,\sum_{i=1}^N
m_i^2.$ Poisson resummation exactly provides the representation of
ordinary Yang-Mills theory on $S^2$ in terms of instantons
\cite{Gross:1994mr,Minahan:1993tp}. Looking closer at
eqs.~(\ref{cipow}) and (\ref{cipow2}) we recognize a similar structure emerging. We
observe the expected exponential of the ``classical action"
$\e^{-\frac{2\pi^2  }{g_s p}\,\sum_{i=1}^N s_i^2}$ and the
fluctuations ${w}_q^{\mathrm{inst}}(s_1,\cdots,s_N)$ which
smoothly reduce to the undeformed ones in the double scaling
limit. The instanton representation is also useful to control the
asymptotic behaviour of the partition function as $g_s\to 0$. In
this limit, only the zero-instanton sector survives, the others
being exponentially suppressed (for fixed $p$).

We immediately recognize that all the non-trivial instanton
contributions are nonperturbative in the $\frac1N$ expansion,
being naively exponentially suppressed, suggesting that the theory
could reduce in this limit to the zero-instanton sector. In order
for this possibility to be correct, one should control the
fluctuation factors. In
ordinary Yang-Mills theory,  the corrections due to
the contribution of instantons to the free energy were calculated
in \cite{Gross:1994mr}. There, it was found that while in the weak-coupling
phase this contribution is exponentially small, it blows up as the
phase transition point is approached. The transition occurs when
the entropy of instantons starts dominating over their Boltzmann
weight $\e^{-S_{\rm inst}}$.

\begin{figure}[htb]
\begin{center}
\epsfxsize=3.2 in \epsfysize=1.8 in \epsfbox{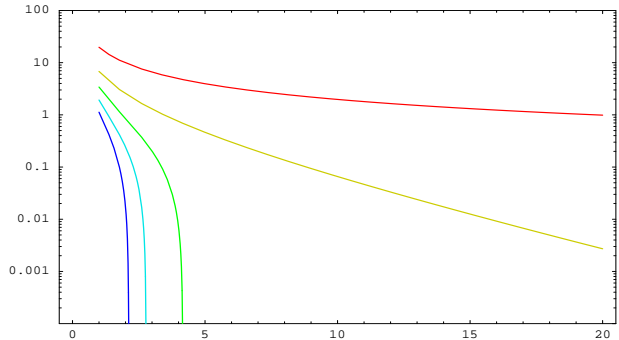}
\end{center}
\caption{\label{PLOT} Plotting $F_0$ for $p=1,2,3,4,5$ as a
function of $t$}
\end{figure}

In principle, the $q$-deformed theory on $S^2$ could experience
the same fate. In eq.~(\ref{cipow}) the Boltzmann weights are the
same as in the undeformed case, and only the structure of the
fluctuations is changed by the deformation. One way to detect the
presence of a phase transition at $N\to\infty$ is to look for a
region in the parameter space where the one-instanton contribution
dominates the zero-instanton sector \cite{Gross:1994mr}. In our
case the ratio of the two contributions is given by (we have
dropped an irrelevant normalization factor)\beq \e^{F_0}=\frac{{
\,\int \prod\limits_{i=1}^N \dd z_i ~\e^{-\frac{8\pi^2 N}{a}\,
\sum_{i=1}^N  z_i^2}\, \prod\limits_{j=2}^N
  \scriptstyle{\left[ \sin^2
  \left(\frac{2\pi\, t  }{a}\,\left( {z_1} - {z_j} \right)\right) -
\sin^2\left(\frac{\pi \,t }{a}\right) \right]}} \,\prod\limits_{i<
j\atop i\ge2} \scriptstyle{\sin^2\left(\frac{2\pi\, t }{a}\,\left(
{z_i} - {z_j} \right)\right)}
  }{{~\e^{\frac{2\pi^2 N
}{a}}\,\int
  \prod\limits_{i=1}^N \dd z_i  ~\e^{-\frac{8\pi^2 N}{a}\,
\sum_{i=1}^N  z_i^2}\, \prod\limits_{1\le i< j\le
N}\scriptstyle{\sin^2\left(\frac{2\pi\, t}a\,\left( {z_i} - {z_j}
\right) \right)}}}
   \ .
\eeq We have employed the saddle-point technique to compute the
above ratio and we recovered the same results of the
one-cut analysis: at $t=t_c$ one-instanton
contributions are no longer suppressed and zero-instanton
approximation breaks down. We do not enter into the details of
this computation but we find instructive to see explicitly how the
behaviour changes above the value $p=2$ by plotting the integral
defining $F_0$  (Fig.~\ref{PLOT}).
The appearing of topological string amplitudes
on the resolved conifold in the large $N$ limit can  be explained as follows. A remarkable
property of the fluctuations is that they do not depend really on
the integers $s_i$ but only their values modulo $p$. It is natural
to organize the partition function by factorizing the independent
fluctuations. We can then write down the partition function in the suggestive form \beq
Z^q_{YM}= \sum_{\{N_k\}}\,\prod_{k=0}^{p-1}{(N_k!)^{-1}\theta_3\left(
\left.\frac{2\pi \ii p}{g_s} \right|
 \frac{2\pi \ii k}{g_s}\right)^{N_k}}Z^p_{\rm CS}\bigl(\{N_k\}\bigr) \
 ,
\eeq where we recognize in the second line the partition function
of $U(N)$ Chern-Simons gauge theory on the Lens space $L_p$ in a
non-trivial vacuum given
by~\cite{Marino:2002fk,Aganagic:2002wv}\beq Z^p_{\rm
CS}\bigl(\{N_k\}\bigr)=\exp\left(-\frac{2\pi^2 }{g_s
p}\,\sum_{m=0}^{p-1} N_m m^2\right)
~w_q^{\mathrm{inst}}(\,\underbrace{0,\dots,0}_{N_0},\dots,
 \underbrace{p-1,\dots,p-1}_{N_{p-1}}\,) \ .\label{LS}\eeq
The relation we have found should be understood as an analytical
continuation to imaginary values of the Chern-Simons coupling
constant $k$ by identifying $g_sp=\frac{2\pi \ii}{k+N}$. The
critical points of the $U(N)$ Chern-Simons action on the manifold
$L_p$ are flat connections which are classified by the embeddings
of the first fundamental group into $U(N)$: one has
$\pi_1(L_p)=\mathbb{Z}_p$. The critical points are therefore given
by discrete $\mathbb{Z}_p$-valued flat connections. They are
easily described by choosing $N$-component vectors with entries
taking values in $\mathbb{Z}_p$. Because the residual Weyl
symmetry $S_N$ of the $U(N)$ gauge group permutes the different
components, the independent choices are in correspondence with the
partitions $\{N_k\}$. The possible vacua of the gauge theory are
in one-to-one correspondence with the choices of flat connections.
The full partition function of Chern-Simons theory involves
summing over all the flat connections, and in fact the exact
answer that can be obtained from the relation with the WZW model
\cite{Rozansky:1993zx} gives such a sum.
Nevertheless, due to the fact that the flat connections here are
isolated points, it is not difficult to extract the particular
contribution of a given vacuum which coincides with
eq.~(\ref{LS}). Let us consider the large $N$ limit of the
zero-instanton sector. Inserting
${w}^{\mathrm{inst}}_q(0,\dots,0)$ into the instanton expansion we
obtain \beq {\cal Z}^{0-{\rm
inst}}=\frac{1}{N!}\,\left(\frac{2\pi}{g_s p}\right)^N
~\e^{-\frac{{g_s}(  N^3-N ) }{6 p}}\,\int
  \prod_{i=1}^N \dd z_i  ~\e^{-\frac{8\pi^2 N}{a}\,
\sum_{i=1}^N  z_i^2} \,\prod_{1\le i< j\le
N}\sin^2\left(\mbox{$\frac{2\pi\, t( {z_i} - {z_j} )} {a}$}\right)
\ . \label{CSM}\eeq The partition function (\ref{CSM}) coincides
with the partition function $Z_{\rm CS}^p$ of Chern-Simons theory
on $L_p$ in the background of the trivial flat-connection with
partition $\{N_k\}=(0,\dots,0)$. According
to~\cite{Marino:2002fk,Aganagic:2002wv}, eq.~(\ref{CSM}) is its
matrix-model representation. The large $N$ limit in the trivial
vacuum  can
be explicitly performed by using the orthogonal
polynomial technique explained in \cite{Marino:2004eq} or it can
be obtained  from the same
result for $S^3$  in \cite{Gopakumar:1998ki} by simply
identifying the parameters. The closed
topological string theory on the resolved conifold ${\cal
O}(1)\oplus{\cal O}(-1)\to \proj^1$ emerges in the limit as an
effect of the geometric transition.

\section{Two-cut solution and the strong-coupling phase}

\begin{figure}[htb]
\centering
\epsfxsize=2.8 in \epsfysize=1.3 in \epsfbox{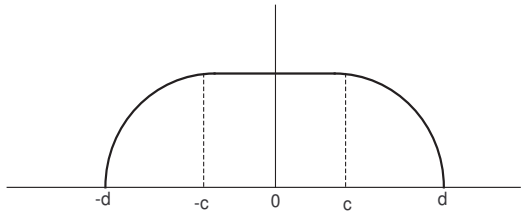}
\caption{\label{dbcut} The double cut ansatz for the distribution
$\rho(z)$.}
\end{figure}

\begin{figure}[htb]
\label{cut1}
\begin{center}
\epsfxsize=3.5 in \epsfysize=1.9 in \epsfbox{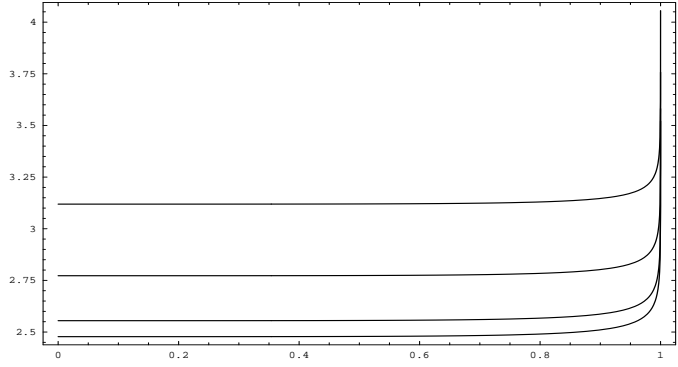}
\end{center}
\caption{\label{saddle} The right-hand side of the saddle-point
  equation times $p$ for $p=3,4,7,20$.}
\end{figure}

For $t>t_c$ we still have to solve the saddle point equation, but
in the presence of the boundary condition $\rho\le 1$. The new
feature which may arise is that a finite fraction of Young
tableaux variables $n_1,\dots,n_N$ condense at the boundary of the
inequality by respecting the parity symmetry of the problem, $
n_{k+1}=n_{k+2}=\cdots=n_{N-k}=0$, while all others are non-zero.
This observation translates into a simple choice for the profile
of $\rho(z)$ as depicted in Fig. \ref{dbcut}.
 In order to respect the bound, the distribution function  is chosen constant
and equal to $1$ everywhere in the interval $[-c,c]$.  In the
intervals $[-d,-c]$ and $[c,d]$ its form is instead dynamically
determined by the saddle-point equation. Let us denote the set
$[-d,-c]\cup [c,d]$ by $\mathcal{{R}}$ and the restriction of the
distribution function to this set by $\tilde \rho=\rho|_{\cal R}$.
The saddle-point equation can be rewritten as \beq
\label{cipow3}\frac{a}{2}\, z-\log\left
|\frac{\sinh\left(\frac{t}{2}\,(z+c)\right)}{\sinh\left(\frac{t}{2}\,(z-c)\right)}\right|=
\frac{t}{2}\,\int_\mathcal{R}\dd w~ {\tilde{\rho}}(w)\, \coth
\left(\mbox{$\frac{t}{2}$}\,(z-w)\right) \ . \eeq Changing
variables $y=\e^{t\, z}~\e^{t^2/a}$ and $u=\e^{t\, w}~\e^{t^2/a}
$, we get: \beq \frac{a}{2 t^2 } \,\frac{\log(y)}{y}-\frac{1}{t\,
y}\,\log \left(\mbox{$\frac{\e^{ c\,t-t^2/a}\, y-1}
   {\e^{-c\,t-t^2/a} \,y -1}$}\right)
= \int_\mathcal{R} {\dd u}~ \frac{\hat\rho(u)}{y-u} \ .
 \eeq We
have defined $\hat \rho(u)\!\!\!\!=\!\!\!\!{\tilde\rho\bigl(\log(u
\e^{-t^2/a})/t\bigr)}/{u t}$ and the support
is now given by ${\cal R}\!=\!\bigl[ \ \underbrace{\e^{-t\,
d+t^2/a}}_{\e^{d_-}}\ \,,\,\underbrace{\e^{-t\,
c+t^2/a}}_{\e^{c_-}}\ \bigr]\cup\bigl[ \ \underbrace{\e^{t
\,c+t^2/a}}_{\e^{c_+}}\ \,,\, \underbrace{\, \e^{t\,
d+t^2/a}}_{\e^{d_+}}\ \bigr] \ . $ The original potential has
been non-trivially modified by a term that depends explicitly on
the endpoints. The saddle point solution is determined by the two
following equations:

\bea &&\frac{a}{2 t}\,\int_0^{\infty} \dd w
~\log(w)\,\frac{\dd}{\dd w}\left(\frac{1}{
\sqrt{(w+\e^{c_-})(w+\e^{c_+})(w+\e^{d_+})(w+\e^{d_-})}}\right)\nonumber\\
&&\qquad\qquad=~\int_{\e^{c_-}}^{\e^{c_+}} \frac{\dd
w}{w}~\frac{1}{\sqrt{(w-\e^{c_-})(\e^{c_+}-w)(\e^{d_+}-w)(w-\e^{d_-})}}
\label{DK1st} \ , \\ &&\frac{a}{2 t}\,\int_0^{\infty} \!\!\!\!\!\!
\frac{\dd w}{
\sqrt{(w+\e^{c_-})(w+\e^{c_+})(w+\e^{d_+})(w+\e^{d_-})}}\nonumber\\
&&\qquad\qquad=~\int_{\e^{c_-}}^{\e^{c_+}}\!\!\!\! \!\frac{\dd
w}{\sqrt{(w-\e^{c_-})(\e^{c_+}-w)(\e^{d_+}-w)(w-\e^{d_-})}} \ .
\label{DK2nd}\eea Eqs. (\ref{DK1st},\ref{DK2nd}) are very
complicated and do not seem promising for an analytical treatment: by
introducing
$k=\sqrt{\frac{(\e^{c_+}-\e^{c_-})(\e^{d_+}-\e^{d_-})}{(\e^{d_+}-\e^{c_-})
(\e^{c_+}-\e^{d_-})}}$
they can be nevertheless expressed in terms of elliptic functions.
The situation then improves dramatically if we change variable
from $k$ to the modular parameter $q=\exp(-\pi\,
K^\prime(k)/K(k))$, $K(k)$ denoting the complete elliptic integral
of the first type. In particular the problem becomes equivalent to
solve a single equation, expressed through an elegant $q$-series
given by \bee \frac{t}{4}=\frac{t_c}{4}-2 p \,\sum_{n=1}^\infty
\frac{(-1)^n}{n}\,\frac{q^{2 n}}{1-q^{2}}\sin^{2}\left(\frac{\pi
n}{p}\right)=-\frac{p}{2}\,
\log\left(\mbox{$\frac{\vartheta_2\left(\left.\frac{\pi}{p}
\right|q\right)}{\vartheta_2\left(\left.0\right|q\right)}$}\right)\equiv \mathfrak{F}(p,q).\label{nos1}
\eeq The graphical behaviour of the function on the right-hand
side of eq.~(\ref{nos1}) is depicted in Fig.~\ref{saddle}, and one
can check that it is a monotonically increasing function as $k$ runs
from $0$ to $1$: $\frac t4$ must be greater than the value of the
right-hand side of eq. (\ref{nos1}) at $k=0$, obtaining the
expected bound $ \displaystyle{ {t}\ge 4\mathfrak{F}(p,0)=
{t_c}} . $
\medskip
 At the critical point $t_c$, we have
$q=0$: the two cuts merge and form a single cut whose endpoints
coincide with those of the one-cut solution. To go further and see
what happens in a neighborhood around $t_c$ we have to solve
eq.~(\ref{nos1}) as a series in $(t-t_c)$. We assume that $q$
admits an expansion of the form $
q=(t-t_c)^\alpha\,\sum_{n=0}^\infty a_n (t-t_c)^n$. Substituting
this expansion  into eq.~(\ref{nos1}), we can iteratively solve
for the coefficients $a_n$ and find (here $\hat t=t-t_c$)\beq
q\!=\!{\sqrt{\hat t}}\left( \frac{\csc (\frac{\pi }{p})}
     {{\sqrt{8 p}}} +
    \frac{{\hat t}\cos (\frac{2\pi }{p})
       {\csc^3 (\frac{\pi }{p})}}{32{\sqrt{2}}
       p^{{3}/{2}}} +
    \frac{{{\hat t}}^2
       \left(   5\cos (\frac{4\pi }{p}) - 32\cos (\frac{2\pi }{p}) -27
        \right)
       {\csc^5 (\frac{\pi }{p})}}{6144{\sqrt{2}}
       p^{{5}/{2}}}+O(\hat t^3)\right)\!\!.
\eeq From this expansion we can obtain all information about the
gauge theory in the strong-coupling phase around the critical
point. To investigate the behaviour of the theory beyond the phase
transition, we need to understand what happens to the distribution
function $\rho$. Because our potential is non-polynomial, we have
no simple relations relating derivatives of the free energy to the
expansion of the resolvent, as in the standard matrix models or in
ordinary Yang-Mills theory. We have to resort therefore to a brute
force calculation. The second
derivative of the free energy can be reduced to computing \beq
\frac{\partial^2{\cal F}}{\partial
a^2}\propto\frac{1}{2}\,\int_{\e^{c_+}}^{\e^{d_+}} \dd
z~\frac{\partial{\rho}(z)}{\partial a}\,
\left(\log(z)-\mbox{$\frac{t^2}a$}\right)^2+\frac{1}{2}\,
\int_{\e^{d_-}}^{\e^{c_-}} \dd z~\frac{\partial{\rho}(z)}{\partial
a}\,\left(\log(z)- \mbox{$\frac{t^2}a$}\right)^2 \eeq as all other
contributions vanish because of the boundary conditions on the
distribution and its symmetries. The derivatives are taken at
constant $t$. A tedious expansion of this quantity around $t=t_c$
using {\sl Mathematica} shows that it vanishes linearly in
$(a-a_c)$ and thus the phase transition is of third order. We can
also present some evidences for how the topological string
expansion emerges. The topological string perturbative expansion
is naturally  organized as a double series in  two modular
parameters $\e^{-t_{s}}$ and $\e^{-{2 t_{s}}/{(p-2)}}$, where
$t_s$ is the K\"ahler modulus (related to our $t$ by $t=\frac{2
t_s}{p-2}$).  The appearance of this double dependence from our
equation is non-trivial and absolutely necessary for the relevant
string interpretation. Because we expect that the topological string
theory will arise when $t$ is large, we have to investigate the
solution of our saddle-point equation around $t=\infty$: $k$ and
consequently $q$ approach $1$ and it is natural to perform a
modular transformation on our equation. This procedure exchanges
$\tau$ and $-\frac1\tau$, and thus a perturbative solution can be
attempted. The saddle-point equation then becomes \bea
\label{nos12} \frac{t}{4} &=&-\frac{p}{2}\,
\log\left(\mbox{$\frac{\vartheta_2\left(\left.
\frac{\pi}{p}\right|q\right)}{\vartheta_2\left(\left.0\right|q\right)}$}
\right)= -\frac{\tilde \tau}{2 p}-\frac{p}{2}\,\log\left(\mbox{$
\frac{\vartheta_4\left(\left.\frac{\ii\tilde \tau}{p}\right|
\tilde q\right)}{\vartheta_4\left(\left.0\right|\tilde q\right)}$}
\right) \eea where we have defined $\tau=\ii K^\prime/K\equiv
\pi/\ii\tilde \tau$ and $\tilde q=\e^{\pi\ii\tilde \tau}$. At
leading order in the solution we have $\tilde \tau= -\frac{t
\,p}{2}=-t_{s}-\frac{2 t_{s}}{p-2}$. The corrections coming from
the theta-functions are exponentially suppressed at this level. To
explore the subleading order we proceed iteratively and we get \beq \tilde
\tau=-t_{s}-\frac{2 t_{s}}{p-2}+4 p^2\,
 \sum_{n=1}^\infty \,\frac{ \e^{-n \,t_{s}-\frac{2 n \,t_{s}}{p-2}}}{1-
 \e^{-2 n\, t_{s}-\frac{4 n\, t_{s}}{p-2}}}\,
\sinh^2\left(\mbox{$\frac{n\, t_{s}}{p-2}$}\right) \ , \eeq and so
on. It is evident that the solution for the modular parameter
$\tilde \tau$ and thus the partition function nicely organizes
into a double expansion in the two moduli $\e^{-t_{s}}$ and
$\e^{-{2t_{s}}/{(p-2)}}$ as expected from string theory.

\subsection*{Acknowledgments}
D.S. warmly thanks the organizers for the kind invitation at the meeting
and the hospitality.
\\

\noindent
This list of references is very far from being complete because of the limited
space and we deeply apologize for that. A complete set of reference can be
found in \cite{Caporaso:2005ta,CGPS}
\\

\newpage
\end{document}